\documentclass[epj]{svjour}
\usepackage{graphics}
\usepackage{amsmath}

\renewcommand{\H}{\hat{H}}
\newcommand{\ed}{\varepsilon_d}
\newcommand{\ep}{\varepsilon_p}
\newcommand{\epx}{\varepsilon_{p_x}}
\newcommand{\epy}{\varepsilon_{p_y}}
\newcommand{\de}[2]{d^\dagger_{#1,#2}}
\newcommand{\dv}[2]{d_{#1,#2}}

\newcommand{\pe}[3]{p^\dagger_{{#1}{#2},#3}}
\newcommand{\pv}[3]{p_{{#1}{#2},#3}}
\newcommand{\ce}[2]{c^\dagger_{#1,#2}}
\newcommand{\cv}[2]{c_{#1,#2}}
\newcommand{\tdp}[3]{t_{#1,{#2}{#3}}}
\newcommand{\tpp}[4]{t'_{{#1}{#2},{#3}{#4}}}
\newcommand{\gk}{\gamma_{\vec{k}}}
\begin{document}
\title{Electronic properties of the three-band Hubbard model}
\author{M. B. Z\"olfl, Th. Maier, Th. Pruschke, J. Keller }                   
\offprints{}          
\institute{Institut f\"ur Theoretische Physik I, Universit\"at
Regensburg, Universit\"atsstr. 31, 93053 Regensburg, Germany}
\date{Received: date / Revised version: date}
%
\abstract{We study the electronic band-structure and transport properties of a 
$\rm CuO_2$-plane within the three-band Hubbard model.
The Dynamical Mean-Field Theory (DMFT) is used to solve the many
particle problem. 
The calculations show that the optical gap $\Delta_{opt}$ is given by
excitations from the lower 
Hubbard band into the so called Zhang-Rice singlet band. 
The optical gap  $\Delta_{opt}$ turns out to 
be considerably smaller than the charge transfer energy $\Delta$ 
($\Delta=\epsilon_p-\epsilon_d$) for a typical set of parameters, 
which is in agreement with experiment.
For the two-dimensional $\rm CuO_2$-plane we investigated the
dependency of the shape of the Fermi surface on the different hopping 
parameters $t_{\rm CuO}$
and $t_{\rm OO}$. A value $t_{\rm OO}/t_{\rm CuO}>0$ leads to a Fermi
surface surrounding the M point.
An additional different static shift of the oxygen energies is also 
considered to calculate the electronic response due to a displacement
of the oxygen atoms given by a frozen phonon.
The density-density correlation for the oxygen orbitals is linear in
doping for both hole and electron doping but shows a different
temperature dependency in the two regimes.
In the first case it is temperature independent and 
increases upon doping, which leads to an increasing electron-phonon
coupling for the $B_{1g}$-mode in high-T$_c$ superconductors. 
} 
\maketitle
\section{Introduction}
\label{intro}
Since the discovery of high temperature superconductivity by Bednorz and 
M\"uller \cite{BedMul} it is a challenging problem to find a
theoretical description of this phenomenon. 
However, even in the normal state the electronic structure of the new 
compounds, which show superconductivity, is extremly hard to describe due to 
electronic correlations. A characteristic feature of the high-T$_c$
compounds is a unit cell with one or more 
$\rm CuO_2$-planes, which are responsible for the superconductivity 
\cite{Dagotto}.
The most simple description of such a $\rm CuO_2$-plane is achieved
within the one-band Hubbard model \cite{Hubbard}, which considers one 
effective 3d orbital of $\rm Cu$ in a tight-binding model in the presence of 
a local Coulomb repulsion $U_d$.
This model describes well the properties of an insulator, because
it reproduces the so called Mott-Hubbard metal-insulator transition 
\cite{MottHubb}. 
However, this simple model leads to a wrong description of the planes concerning 
the doping dependency of various properties, such as the asymmetric
magnetic doping-temperature phase diagram \cite{Almasan}.        
  
The oxygen atoms in the $\rm CuO_2$-plane, respectively the $2p$-orbitals of these 
atoms introduce a further degree of freedom and 
the strongest hybridization takes place between the 
$\rm Cu$-$3d_{x^2-y^2}$ orbital and the $\rm O$-$2p_{x/y}$ orbitals 
\cite{Fulde}. Therefore they form the lowest lying bonding state and
the highest lying anti-bonding state. Choosing the most simple
description of the $\rm CuO_2$-plane one is led to the hole
picture and one has to consider the latter state, 
which is occupied with one hole.
In order to describe the additional degrees of 
freedom, one has to solve at least a so called three-band Hubbard
model or Emery model \cite{Emery}, with two additional
oxygen orbitals 2$p_x$ and 2$p_y$, 
which are included in a tight-binding manner. For the sake of simplicity all
Coulomb energies but $U_d$ are neglected.
Within this model one is able to describe a doped charge transfer insulator, 
which gives a more realistic picture of the planes in a high-T$_c$ compound.
The three-band model has recently been studied by Schmalian et al. 
\cite{Schmalian} using a generalized dynamical mean-field (DMFT)
approach and our contribution is based on their approach.
Watanabe et al. \cite{Watanabe} discussed the metal-insulator
transition in a two-band Hubbard model in infinite dimensions and 
discovered the coexistence of
metallic and antiferromagnetic phases. This was also found by Maier
et al. \cite{Maier}, who additionally reported about an asymmetric 
magnetic doping-temperature phase diagram within our approach, which
shows, that the antiferromagnetic phase is more stable upon electron
doping than upon hole doping.

In this paper the DMFT approach of Schmalian et al. \cite{Schmalian}
for the three-band Hubbard model is used. 
Several extra features in two spatial dimensions were added, such as 
an oxygen-oxygen hopping process and a splitting of the 
oxygen energy levels. 
We discuss results for the one-particle spectra 
of the orbitals and for the in-plane conductivity for hole and
electron doped systems considering a semi-elliptic density of states. 
We further discuss, for the case of a two dimensional system, how an 
additional oxygen-oxygen hopping process affects the shape of the 
Fermi surface.
Finally a static shift of the oxygen energies due to a displacement given 
by a frozen phonon is considered in order to calculate the static 
density-density correlation, which describes the electronic response
to this kind of distortion.
A summary will conclude the paper.  
      
\section{The three-band Hubbard model}
\label{sec:1}
\begin{figure}
\centering{\resizebox{0.35\textwidth}{!}{\includegraphics{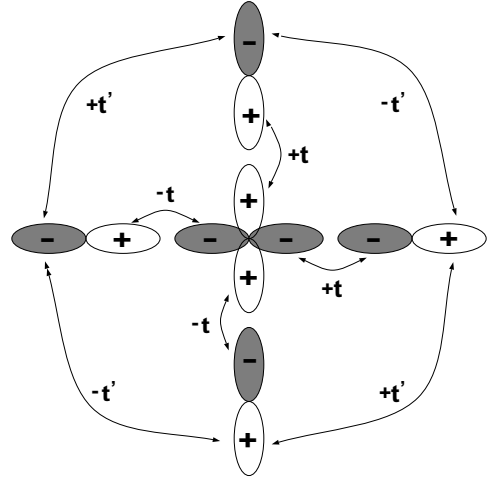}}}
\caption{Used phase convention, which determines the phases of all hopping 
processes.}
\label{fig:1}    
\end{figure}

The starting point of our approach is a simplified Emery Hamiltonian 
\cite{Emery} describing the dynamics of holes in a doped 
$\rm Cu O_2$-plane. 
The nearest neighbour hopping processes between
the 3d-orbital of the $\rm Cu$-atom and the 2$p_x$-/ 2$p_y$-orbital of the 
$\rm O$-atom ($\tdp{i}{j}{\nu}$ and $\tpp{i}{\kappa}{j}{\nu}$) are taken into 
account. $\ed$, $\epx$ and $\epy$ represent the energy levels of each
orbital and $\mu$ the chemical potential, in addition a local Coulomb
energy $U_d$ for the case of a doubly
occupied d-orbital, which is responsible for the correlations.    
\begin{equation}
\begin{split}
  \H &=\sum_{i,\sigma} (\ed-\mu)\:\de{i}{\sigma} \dv{i}{\sigma} 
+ \sum_{i,\nu,\sigma} ({\ep}_\nu-\mu)\:\pe{i}{\nu}{\sigma}\pv{i}{\nu}{\sigma}\\
   + &\sum_{i,j,\nu,\sigma}(\tdp{i}{j}{\nu}\: 
        \de{i}{\sigma} \pv{j}{\nu}{\sigma}+h.c.) 
+ \underset{i\ne j,\nu\ne\kappa }{\sum_{i,\nu,j,\kappa,\sigma}}
        \tpp{i}{\kappa}{j}{\nu}\: 
        \pe{i}{\kappa}{\sigma}\pv{j}{\nu}{\sigma}                          \\
   + &\sum_{i} U_d \:\de{i}{\uparrow}\dv{i}{\uparrow}
            \de{i}{\downarrow}\dv{i}{\downarrow}\:.
\end{split}
\end{equation}
As mentioned earlier, this Hamiltonian is able to describe a doped
charge transfer insulator, which 
is characterized  by the charge transfer energy $\Delta=\ep-\ed$.
The relation $U>\Delta$ ensures the system of being in the 
charge transfer regime \cite{Zaanen}. This relation typically holds for 
parameters obtained from a first principles calculation (see table
\ref{tab:1}).     
\begin{table}[h!]
\begin{center}
\caption{Parameters for a three-band model (in eV) calculated  with 
a constrained first principles calculation for $\rm La_2CuO_4$ done by
Hybertsen et al. \cite{Hybertsen}.}
\label{tab:1}     
\begin{tabular}{lllllll}
\hline\noalign{\smallskip}
$\Delta$ & $t$ & $t^\prime$ & $U_d$ & $U_p$ & $U_{pd}$ & $U_{pp}$ \\
\noalign{\smallskip}\hline\noalign{\smallskip}
3.6 & 1.3  & 0.65  & 10.5  & 4 & 1.2  & 0  \\
\noalign{\smallskip}\hline
\end{tabular}
\end{center}
\end{table}
The gauge invariance of the Hamiltonian allows us to choose the phase of the 
wave functions freely. We use the phase convention shown in figure
\ref{fig:1}, which determines all phases of the considered hopping processes. 
The transformation into $\vec{k}$-space leads us to the following
Hamiltonian (for simplicity we show the non-interacting part only):
\begin{equation}
H_o=\sum\limits_{\vec{k},\sigma} \:
(d^{\dagger}_{\vec{k}\sigma},p^{\dagger}_{x\vec{k}\sigma}
,p^{\dagger}_{y\vec{k}\sigma}) \:\underline{\underline{h}}(\vec{k})\:
\begin{pmatrix}
d_{\vec{k}\sigma}\\p_{x\vec{k}\sigma}\\
p_{y\vec{k}\sigma}
\end{pmatrix},
\end{equation}
with the matrix $\underline{\underline{h}}(\vec{k})$:
\begin{equation}
\small
\begin{pmatrix} 
\ed-\mu & 2it\sin(\vec{k}\vec{r_x}) & 2it\sin(\vec{k}\vec{r_y}) \\
-2it\sin(\vec{k}\vec{r_x})&\varepsilon_{p_x}-\mu&
-4t^\prime\sin(\vec{k}\vec{r_x})\sin(\vec{k}\vec{r_y})\\
-2it\sin(\vec{k}\vec{r_y}) & -4t^\prime\sin(\vec{k}\vec{r_x})\sin(\vec{k}\vec{r_y})&\varepsilon_{p_y}-\mu
\end{pmatrix}.
\end{equation}
This uncorrelated Hamiltonian can be diagonalized easily. In order to
treat the strong Coulomb repulsion $U_d$, we use the dynamical
mean-field theory, where the lattice problem is
mapped onto an effective Anderson impurity model \cite{Anderson}. This model 
describes an impurity and its hybridizations with conduction electrons
(holes in our case). 
The mapping onto this model is exact in infinite dimensions \cite{Metzner}
and means the restriction to local selfenergy diagrams in finite dimensions.   

\noindent
As first step of our derivation of the cluster/impurity we apply a
unitary transformation to the Hamiltonian (2):  
\begin{equation}
H=\sum\limits_{\vec{k},\sigma} 
(d^{\dagger}_{\vec{k}\sigma},p^{\dagger}_{\vec{k}\sigma}
,{\bar{p}}^{\dagger}_{\vec{k}\sigma}) 
\begin{pmatrix}
  \ed-\mu  & - 2 t \gk                  & 0                           \\
 - 2 t \gk & \varepsilon_{p\vec{k}}-\mu & t^\prime_{\vec{k}}          \\
  0        & t^\prime_{\vec{k}}         & \varepsilon_{\bar{p}\vec{k}}-\mu
\end{pmatrix}
\begin{pmatrix}
d_{\vec{k}\sigma}\\p_{\vec{k}\sigma}\\
{\bar{p}}_{\vec{k}\sigma}
\end{pmatrix}
\end{equation}
with
\begin{align}
\small
\varepsilon_{p\vec{k}}=\frac{1}{\gk^2}\{\sin(\vec{k}\vec{r_x})^2& \epx
+\sin(\vec{k}\vec{r_y})^2\epy-\nonumber\\
&-8t^\prime\sin(\vec{k}\vec{r_x})^2
\sin(\vec{k}\vec{r_y})^2\},\\
\varepsilon_{\bar{p}\vec{k}}=\frac{1}{\gk^2}\{\sin(\vec{k}\vec{r_y})^2&\epx
+\sin(\vec{k}\vec{r_x})^2\epy+\nonumber\\
&+8t^\prime\sin(\vec{k}\vec{r_x})^2
\sin(\vec{k}\vec{r_y})^2\},\\
t^\prime_{\vec{k}}=\frac{\sin(\vec{k}\vec{r_x})\sin(\vec{k}\vec{r_y})}{\gk^2}
&\times\nonumber\\
&\hspace{-2cm}
\{(\epx-\epy)+4t^\prime(\sin(\vec{k}\vec{r_x})^2-\sin(\vec{k}\vec{r_y})^2)\}
\end{align}
and
\begin{equation}
\gk^2=\sin(\vec{k}\vec{r_x})^2+\sin(\vec{k}\vec{r_y})^2.
\end{equation}
This transformation leads to new orbitals
$p$ and $\bar{p}$. $d$ hybridizes with $p$ but not with $\bar{p}$.
The case of $\epx=\epy=\ep$ and $t^\prime=0$ leads to a
dispersionless $\bar{p}$-band at $\ep-\mu$ and the 
$\bar{p}$-orbital is decoupled from the rest of the system. 
This procedure leads to a well-defined description of local orbitals 
at different lattice cells, which was suggested by Valenti and Gros 
\cite{Valenti}. 
Within this formulation the problem how
to distribute the $p$-orbitals, which orginally are located in between 
different $Cu$-sites, to a particular unit cell was solved.
In order to construct a sensible DMFT we here follow a
path, which identifies the effective 'impurity' needed to set up the
DMFT equations directly from the d-part of the Green's function.
Due to the form  (4) of the Hamiltonian the general structure for 
$G_{\vec{k}\sigma}^{dd}(z)$ within the DMFT is:
\begin{equation}
\begin{split}
G&^{dd}_{\vec{k}\sigma}(z)= \\
&\left[ z -\ed +\mu -\Sigma_d(z)-\frac{4t^2\gk^2}
{z-\varepsilon_{p\vec{k}}+\mu-\frac{\displaystyle{t^\prime_\vec{k}}^2}
              {\displaystyle z-\varepsilon_{\bar{p}\vec{k}}+\mu} } 
   \right]^{-1}\:\:.
\end{split}
\end{equation} 
The sum over $\vec{k}$ will in general lead to a complicated structure
in the denominator. However, since the selfenergy $\Sigma_d(z)$ is
$\vec{k}$-independent we may cast it into the form:
\begin{equation}
\begin{split}
G&^{dd}_{\sigma}(z)=\frac{1}{N}\sum_{\vec{k}} G^{dd}_{\vec{k}\sigma}(z)\overset{!}{=}\\ 
&\left[ z -\ed +\mu -\Sigma_d(z)-\frac{{t_{00}}^2}
{z-\varepsilon_{p00}+\mu-\Delta(z)} \right]^{-1}\:\:,
\end{split}
\end{equation} 
where $t_{00}$ and $\varepsilon_{p00}$ are Fourier-transformed terms  
of $-2t\gk$ and $\varepsilon_{p\vec{k}}$. 
The effective medium's function $\Delta(z)$ is defined through
equation (10) and incorporates  all bandstructure effects  due to the
dispersion and hybridizations. Note that this definition is far from
being unique! In fact we may choose for example 
\begin{equation}
\frac{1}{N}\sum_{\vec{k}} G^{dd}_{\vec{k}\sigma}(z)\overset{!}{=} 
[ z -\ed +\mu -\Sigma_d(z)-\tilde{\Delta}(z) ]^{-1}
\end{equation} 
as another possibility. In this case the equation defining the DMFT
would be formally the same as for the one-band Hubbard model. However, 
due to the singular  structure  induced by a d-p-hopping equation (10)
turns out to be numerically more convenient: For $\Delta(z)=0$ the
level structure of the d-p complex is already included from the 
outset and $\Delta(z)$ is rather smooth. On the other hand the at the
first glance more natural choice (11) would result in a
$\tilde{\Delta}(z)$ with a highly singular behaviour   
reflecting the existence of unrenormalized p-states. 
Thus the Hamiltonian of the effective impurity is defined by:
\begin{equation}
\begin{split}
H  &=  \sum_{\sigma} (\ed-\mu)\:d^{\dagger}_{\sigma} d_{\sigma} \quad
 +  \sum_{\sigma} ({\varepsilon_{p}}_{00}-\mu)\:p^{\dagger}_{\sigma} p_{\sigma}\\
&+  \sum_{\sigma} (t_{00}\:d^{\dagger}_{\sigma} p_{\sigma} + h.c.)  
+  U_d\: d^{\dagger}_{\uparrow} d_{\uparrow} 
d^{\dagger}_{\downarrow} d_{\downarrow}\\
&+ \sum_{\vec{k},\sigma} (V_{\vec{k}}\:\,
  p^\dagger_{\sigma}\cv{\vec{k}}{\sigma} + h.c.)
 + \sum_{\vec{k},\sigma}
  \varepsilon_{\vec{k}}\:\ce{\vec{k}}{\sigma}\cv{\vec{k}}{\sigma}\:, 
\end{split}
\end{equation}
with a hybridization $V_{\vec{k}}$ of the $p$-orbitals to the
effective medium described by the dispersion $\varepsilon_{\vec{k}}$.
The effective medium enters the solution of the impurity problem only
by the hybridization function 
\begin{equation}
\Delta(z)=\frac{1}{N}\sum_\vec{k}
\frac{|V_{\vec{k}}|^2}{z-\varepsilon_{\vec{k}}}\:\:.
\end{equation}
of the $p$-states, which has to be determined selfconsistently
from equation (10).

Thus equation (10) is the central point of our selfconsistency scheme.
After calculating the local Green's function by the sum of $\vec{k}$
one can extract the hybridization function $\Delta(z)$, 
which is used to solve the impurity problem with help of an extended
NCA scheme \cite{Schmalian}. 
From the impurity Green's function for the $d$-orbital we extract the
selfenergy $\Sigma(z)$, which is then used in the calculation of the
lattice Green's function in equation (10). 
 

\section{The effective two-band model}
\label{sec:3}

Let us begin the discussion of the properties of the three-band
Hubbard model by considering the standard case $\ep=\epx=\epy$ and
$t^\prime=0$. Then the non-hybridizing $\bar{p}$-orbital decouples
from the rest of the system leading to a two-band problem. This problem 
can in principle be generalized to D dimensions, but it is impossible 
to find a uniform scaling to obtain a nontrivial limit $D\to\infty$ 
\cite{Valenti}. 
The DMFT used here therefore always leads to an approximate treatment
of the model.
In order to simplify numerical calculations the semi-elliptic density
of states is used, i.e.:

\begin{equation}
\begin{split}
G_d(z) &=\frac{1}{N}
\sum\limits_{\vec{k}}
\frac{1}
{z-(\ed-\mu)-\Sigma(z)-\frac{\displaystyle 4t^2\gk^2}
{\displaystyle z-(\ep-\mu)}}=\\
       &=\int d\varepsilon
    \frac{\rho_o(\varepsilon)}
     {z-(\ed-\mu)-\Sigma(z)-
        \frac{\displaystyle 4t^2(1-\varepsilon)}
             {\displaystyle z-(\ep-\mu)}}\:,
\end{split}
\end{equation}
with $\rho_o(\varepsilon)=\frac{2}{\pi}\sqrt{1-\varepsilon^2}$.
The Green's function for the $p$-orbital is obtained, if one exchanges
the energies $\ed+\Sigma(z)$ and $\ep$.

\subsection{One-particle spectra}
\label{sec:4}
\begin{figure}[ht!]
\centering{\resizebox{0.35\textwidth}{!}{\includegraphics{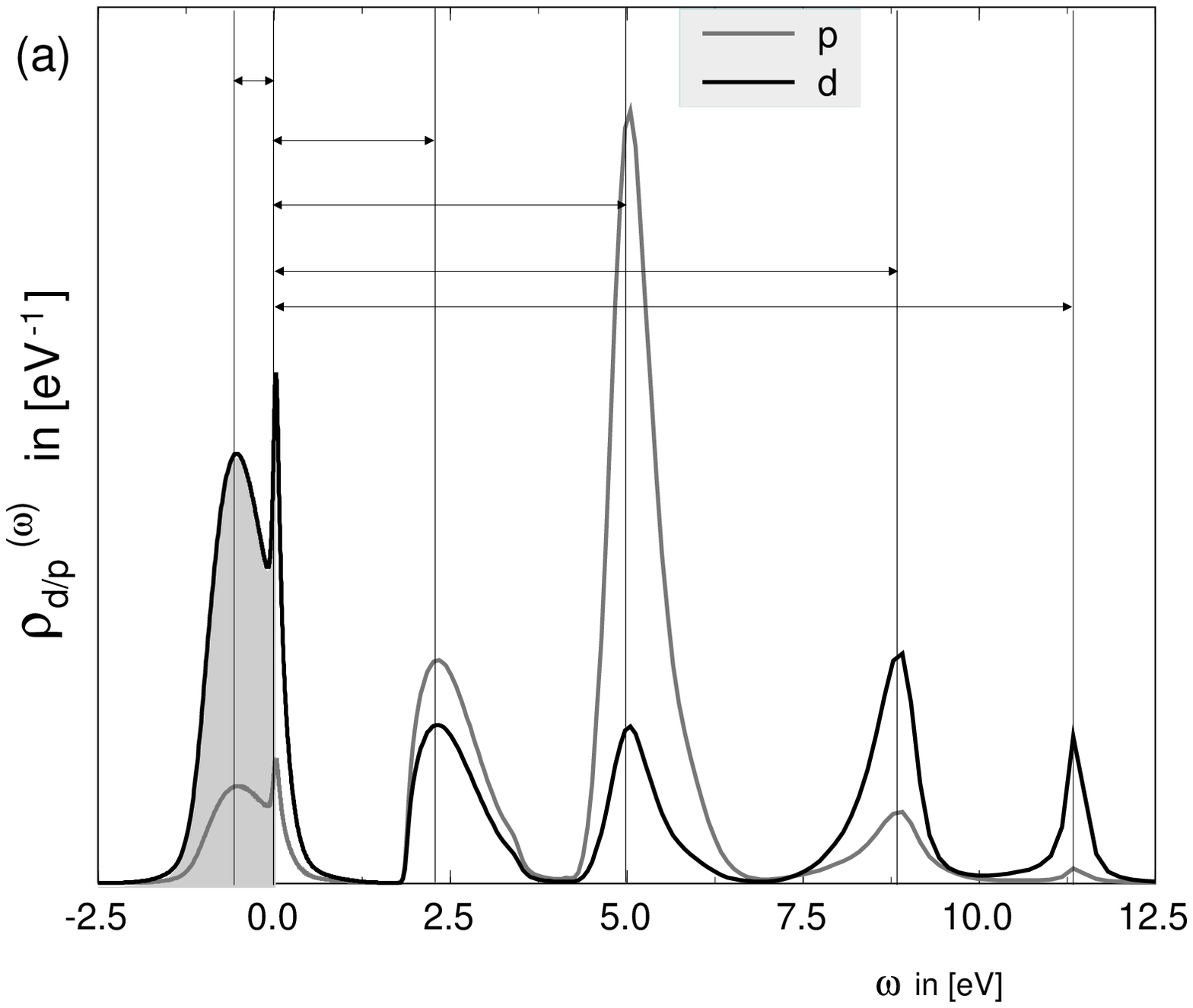}}}\\
\vspace{1.5cm}
\centering{\resizebox{0.35\textwidth}{!}{\includegraphics{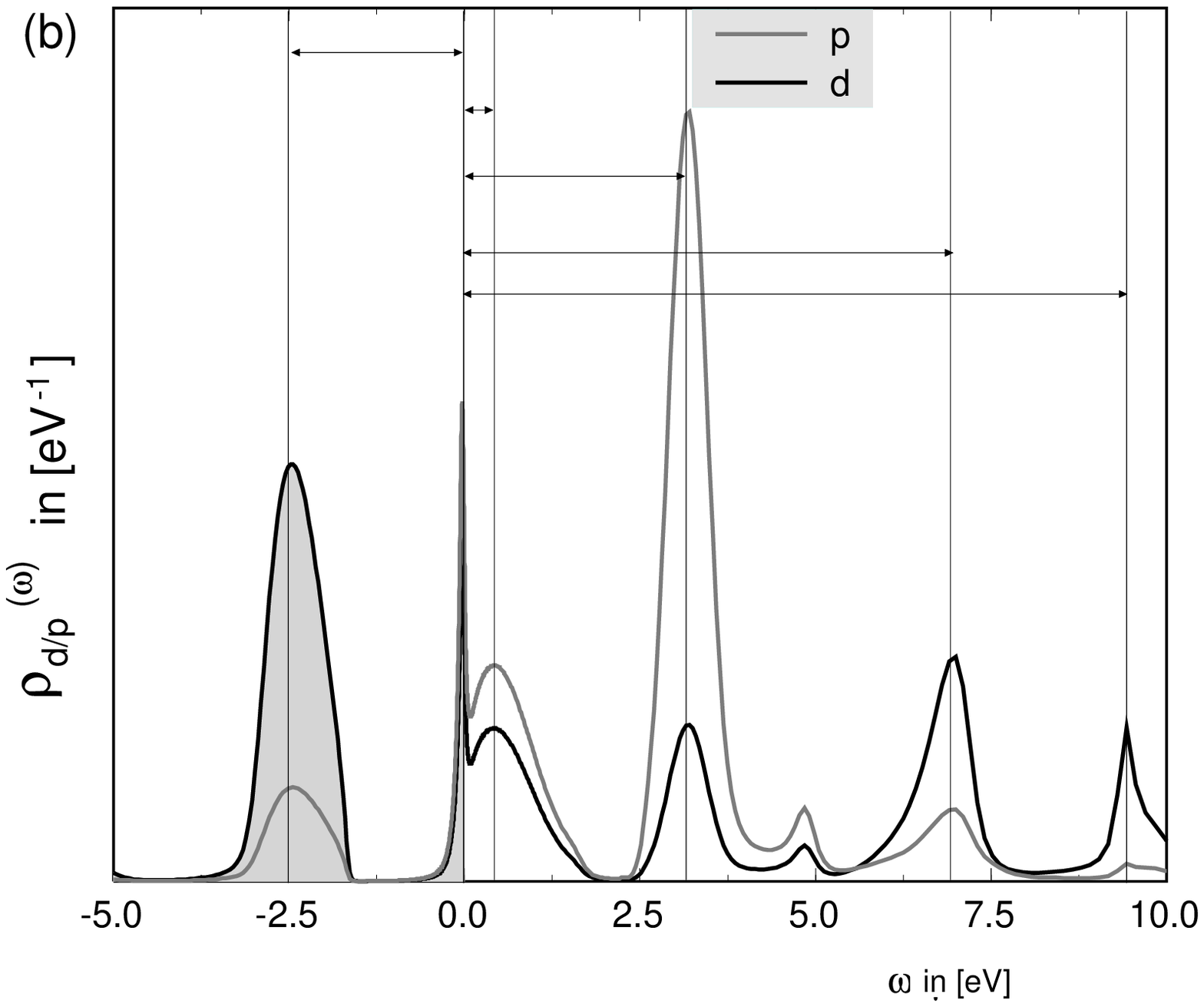}}}
\caption{One-patricle spectra for the d-orbital and the hybridizing 
p-orbital in a electron doped system (a) and in a hole doped  
system (b) for energies in units of the hoppping parameter t. 
The parameter set was $\beta=40.0\:\:eV^{-1}$, $\Delta=3.6\:\:eV$,
$t=1.0\:\:eV$ and $U_d=7.2\:\:eV$.}
\label{fig:2}    
\end{figure}

Figures \ref{fig:2}a and \ref{fig:2}b show the one-particle spectra for the 
$d$-orbital and the hybridizing $p$-orbital for an electron and a hole
doped system, respectively,  at room temperature, i.e. at a reciprocal
temperature of $\beta=40\:\:eV^{-1}$.
They show in terms of the hole picture from the left to the right: 
the lower Hubbard band, the Zhang-Rice band mostly due to exitations
into a two particle singlet state, 
the $p$-band and the upper Hubbard band, which has a double  
peak structure due to the complex excitation capabilities of the
cluster under consideration. 
In both figures typical particle hole excitations are marked
by arrows, which play a role in the dynamical conductivity 
(see section \ref{sec:5}).
In addition to those expected structures a sharp 
resonance appears near the Fermi energy in both doping regimes. 
From the DMFT of the one-band Hubbard model it is well known \cite{prukot},
that this resonance is connected to a local Kondo-like screening. It thus is 
frequently termed Abrikosov-Suhl resonance to stress the similarity of
its physics to the Kondo effect. 
For a detailed description of the doping and temperature 
dependencies of the spectral weigths see ref. \cite{Schmalian}.      

\subsection{The in-plane conductivity}
\label{sec:5}
In order to calculate the in-plane conductivity in linear response theory 
using Kubo's formula, one has to derive the current-current correlation 
function. The current density operator for the three-band model
( see equation (1)) is given by the time derivative 
of the polarization operator using Heisenberg's equation of motion.
This procedure leads to the expression:
\begin{equation*}
\vec{j} = -\frac{2et}{\hbar}\sum\limits_{\nu} 
          \sum\limits_{\vec{k},\sigma} \vec{r}_\nu \sin(\vec{k}\vec{r}_\nu)
          (\de{\vec{k}}{\sigma}p_{\nu\vec{k},\sigma}
           +p^\dagger_{\nu\vec{k},\sigma}\dv{\vec{k}}{\sigma}) ,
\end{equation*}
with the vector $\vec{r}_\nu$ connecting $d$- and $p$-sites.

In the following the real part of the in-plane conductivity in x-direction is examined. 
The current-current correlation function is
calculated within the orbital representation of equation (4).
Since the particle-hole excitations with the non-hybridizing $\bar{p}$-states are  
playing a role only at high energies, contributions with $\bar{p}$-orbitals
are neglected. Thus we obtain:      
\begin{equation*}
\begin{split}
&\Re\left\{\sigma_x(\omega)\right\}= \frac{1}{2}
(\Re \left\{\sigma_x(\omega)\right\}+\Re
\left\{\sigma_y(\omega)\right\})
=\\
&=\frac{1}{2} (\frac{aet}{\hbar})^2  \frac{1}{N}\sum_\vec{k} \frac{\sin(\vec{k}\vec{r}_x)^4+\sin(\vec{k}\vec{r}_y)^4}{\gk^2}
\times\\
&\quad\quad\quad\quad\int_{-\infty}^\infty d\varepsilon 
\frac{f(\varepsilon)-f(\varepsilon+\omega)}{\omega} \:\: 
\Pi(\varepsilon,\omega,\vec{k})
\end{split}
\end{equation*}
with 
\begin{equation*}
\Pi(\varepsilon,\omega,\vec{k})=\sum_{(\alpha,\beta)}
\rho_{\alpha}(\varepsilon+\omega,\vec{k}) 
\rho_{\beta}(\varepsilon,\vec{k})\:\:
\end{equation*}
and spectral function $\rho_\alpha$ of Green's functions with the following orbital
combinations:
\begin{equation*} 
(\alpha,\beta)\:\epsilon\:
\{(pp^\dagger,dd^\dagger),(dd^\dagger,pp^\dagger),(pd^\dagger,pd^\dagger),
  (dp^\dagger,dp^\dagger)\}\:.
\end{equation*} 
Due to the locality of the vertex kernel 
and the antisymmetric current operator vertex corrections 
vanish analogously to  calculations for the one-band Hubbard model 
\cite{LeitPru}. 
Finally the $\vec{k}$-summation is carried out in an averaged way by
integration over the pure density of states:
\begin{equation*}
\begin{split}
&\Re\left\{\sigma(\omega)\right\}=\\
&=\sigma_o 
\int_{-\infty}^\infty\int_{-\infty}^\infty
d\varepsilon d\tilde{\varepsilon} \:\:
\rho_o(\tilde{\varepsilon}) \:\:  
\frac{f(\varepsilon)-f(\varepsilon+\omega)}{\omega} \:\: 
\Pi(\varepsilon,\omega,\tilde{\varepsilon}).
\end{split}
\end{equation*}

\begin{figure}[ht!]
\centering{\resizebox{0.4\textwidth}{!}{\includegraphics{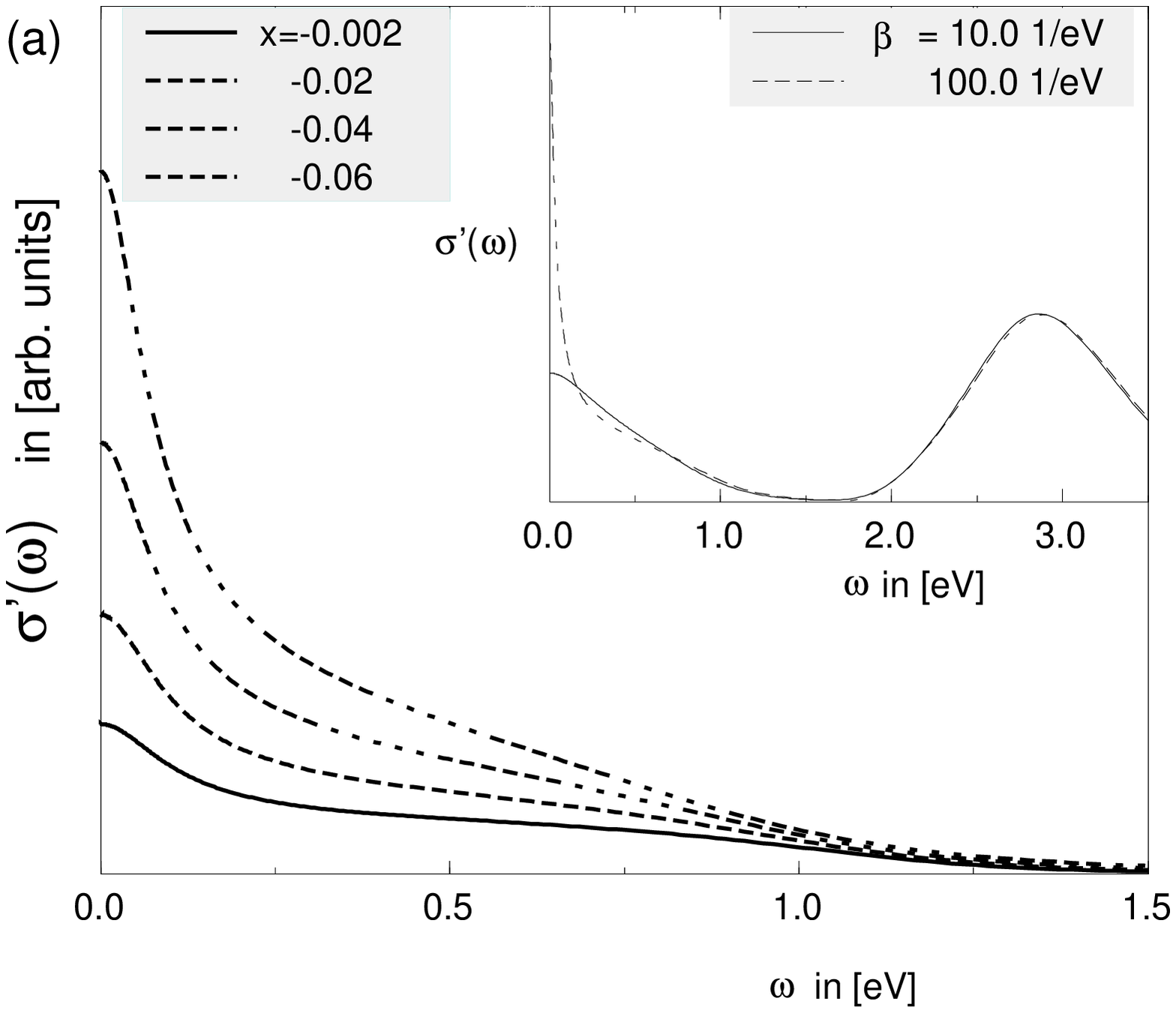}}}\\ 
\vspace{0.5cm}
\centering{\resizebox{0.4\textwidth}{!}{\includegraphics{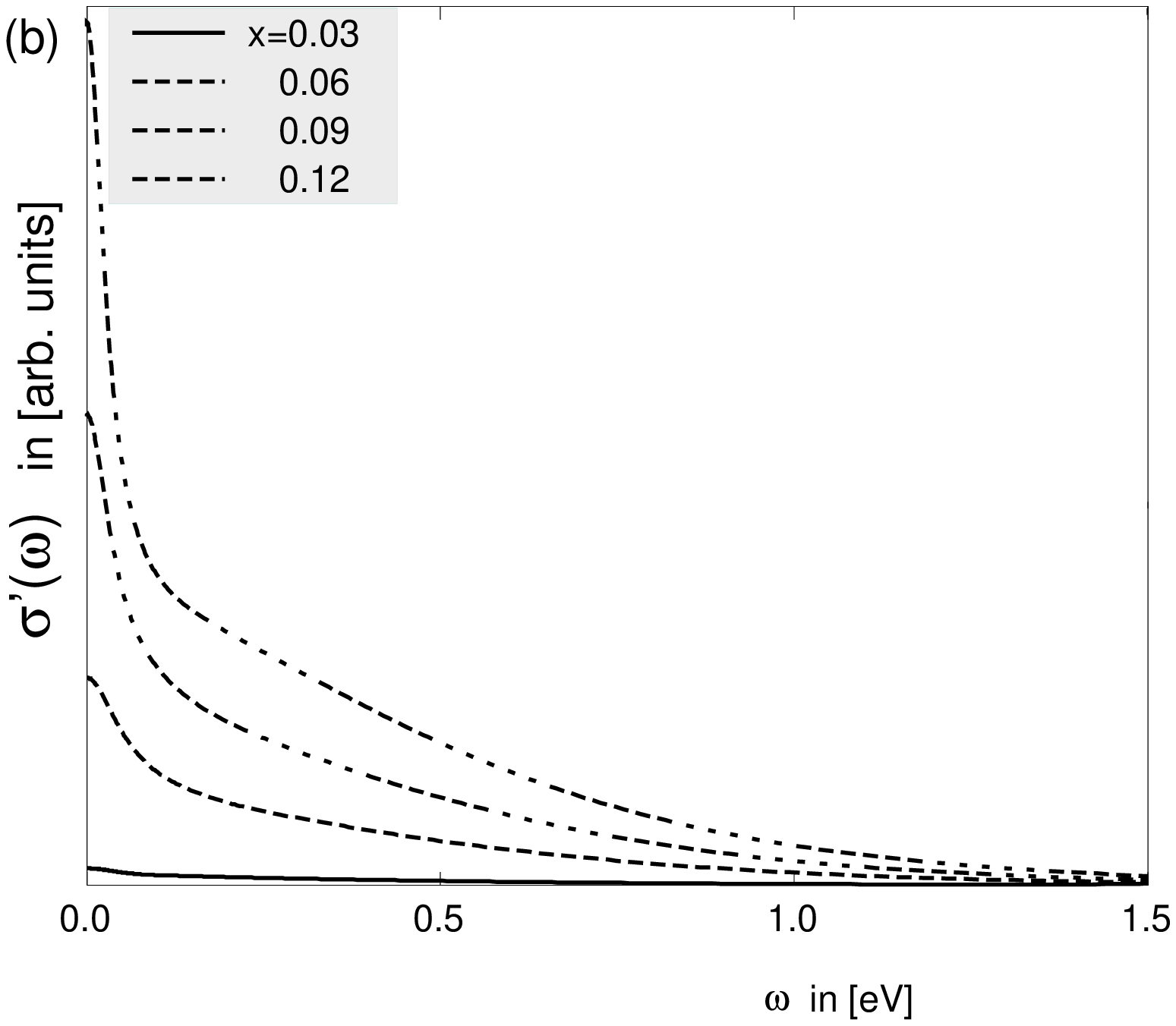}}}
\caption{Real part of the conductivity for several electron doped (a)
and several hole doped systems (b) at $\beta=40\:eV^{-1}$. 
The inset shows the temperature dependency for a
system with the doping level $x=-4\:\%$.}
\label{fig:4}    
\end{figure}

Typical particle-hole excitations for electron and
hole doped systems are denoted by the vertical arrows in figure
\ref{fig:2}a and \ref{fig:2}b. 
These different exictations can in principle all be resolved in 
the graph of the conductivity. 
However, for sake of simplicity only the low energy scale is shown in figures 
\ref{fig:4}a and \ref{fig:4}b.
The in-plane conductivity was calculated for several doping parameters $x$ for 
a reciprocal temperature $\beta=40\:\:eV^{-1}$. 
The temperature dependency for constant doping x is shown in the
inset of the figure \ref{fig:4}a.
Both electron and hole doped systems show an increasing conductivity
upon doping in the mid-infrared region at $0.5\:eV$. The far-infrared part 
of the conductivity shows a pronounced increase due to the growth of
spectral weight of the Abrikosov-Suhl resonance upon doping.
In the inset of figure \ref{fig:4}a the strong temperature dependency of the 
low energy excitations is shown.
The broad peak at $3\:eV$ shown in the inset corresponds to transitions between 
the lower Hubbard band and the Zhang-Rice band.
The energy separation of these two bands, which describes the optic energy gap 
$\Delta_{opt}$, is smaller than the bare charge transfer energy $\Delta=\ep-\ed$.
For example, for the present set of parameters we find 
$\Delta_{opt}\approx2\:eV$, which is in good agreement with meassurements by Uchida 
et al. \cite{Uchida} for $\rm La_{2-x}Sr_x CuO_4$ and $\rm Nd_{2-x}Ce_x CuO_{4-y}$ .

\section{Results for a quadratic $\rm CuO_2$-system}
\label{sec:6}

In this section we present results for a two-dimensional tight-binding model. 
At a first glance the use of the DMFT as an approximation may seem to
be particularly crude. However,
as long as the system is not too
close to a phase transition the choice of a local selfenergy
appears to capture the most important dynamics, which is seen if one 
compares the results from a local approximation with those of a Monte Carlo
calculation in the two dimensional one-band Hubbard model \cite{PruMon}.

\subsection{Bandstructure and Fermi surface}
\label{sec:7}

\begin{figure}[ht!]
\centering{\resizebox{0.45\textwidth}{!}{\includegraphics{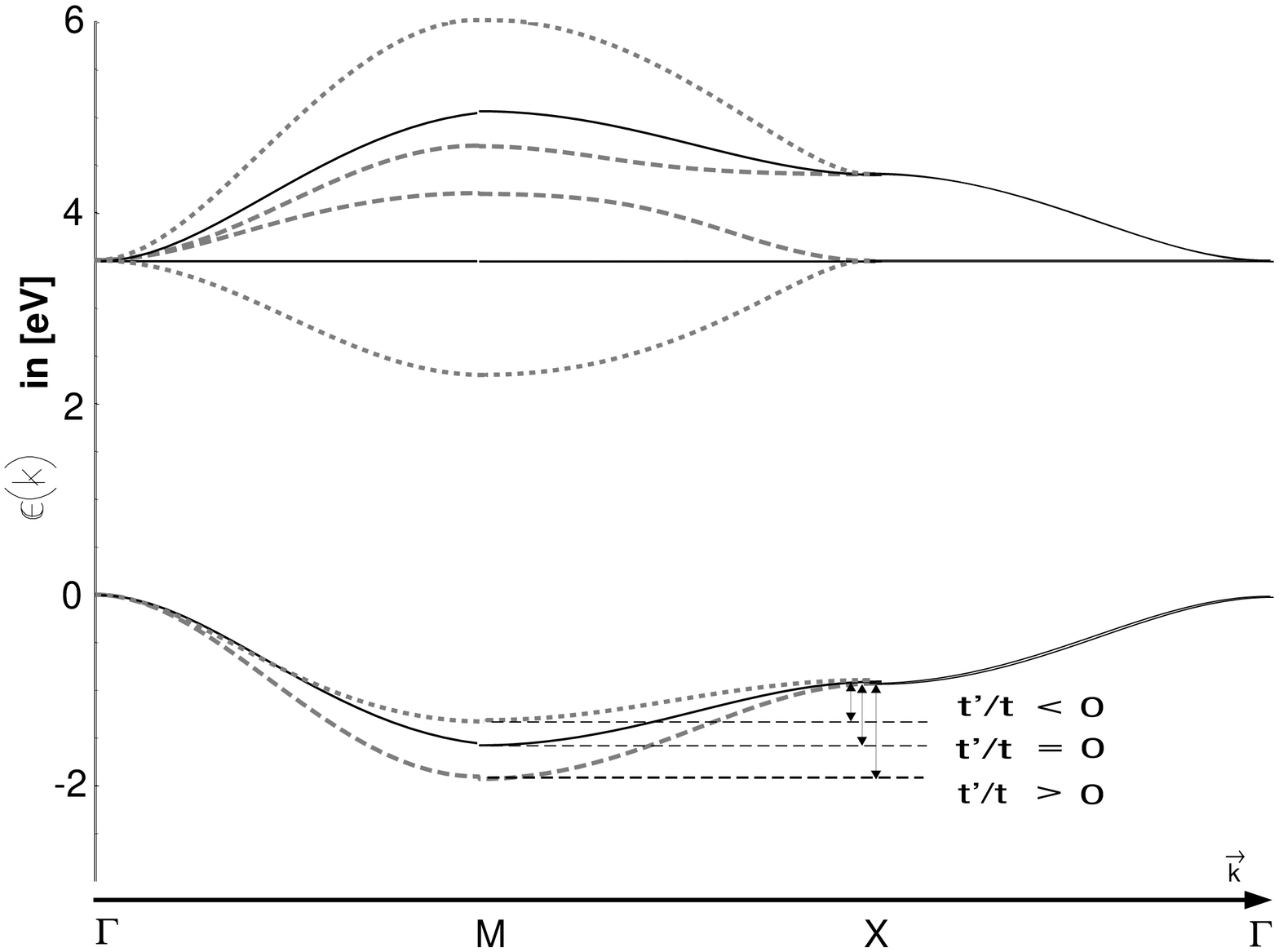}}}
\caption{Tight-binding bandstructure for several values
of $t^\prime/t$. $\ed=0\:eV$, $\ep=3.6\:eV$, $t=1\:eV$ and
$t^\prime=0,\pm0.3\:eV$.}
\label{fig:6}    
\end{figure}

Before we discuss the bandstructure for the correlated electron system, 
let us look at the uncorrelated tight-binding bandstructure, which is shown 
in figure \ref{fig:6}.
In the following the name of the bands is given by their main orbital
character.
Here the band with the lowest energy (hole picture) has primarily 
$d$-character. The upper bands have $p$-character.
For hole-doping the Fermi energy lies in the upper bands, for
electron-doping in the lower band.
The $d$-band has a minimum at the $M$-point and a saddle point at
the $X$-point. With increasing value of $t^\prime/t$ the distance
between the minimum and the saddle point increases. This means that the
region of electron doping, where one can find 
a Fermi surface surrounding the $M$-point by adjusting the chemical 
potential $\mu$, is also increasing.
In the hole doped case, however, a positive value of $t^\prime/t$ leads
to a Fermi surface centered around $\Gamma$ produced by the
hybridizing $p$-band and one around $M$ by the non-hybridizing $\bar{p}$-band, 
since the non-hybridizing $p$-band is pushed towards  the hybridizing $p$-band 
with an increasing value of $t^\prime/t$.
On the other hand, a value $t^\prime/t<0$ would
push down the non-hybridizing $\bar{p}$-band, leading to a dispersion
minimum at the $M$-point, as it is observed experimentally \cite{rep}.
Thus the requirement of a Fermi surface surrounding the $M$-point for both
electron and hole doping means that one 
either has to assume different signs for $t^\prime$ in the two cases of
doping or look for a more subtle
mechanism leading to the observed physics for a fixed sign of $t^\prime$.

\begin{figure}[ht!]
\centering{\resizebox{0.4\textwidth}{!}{\includegraphics{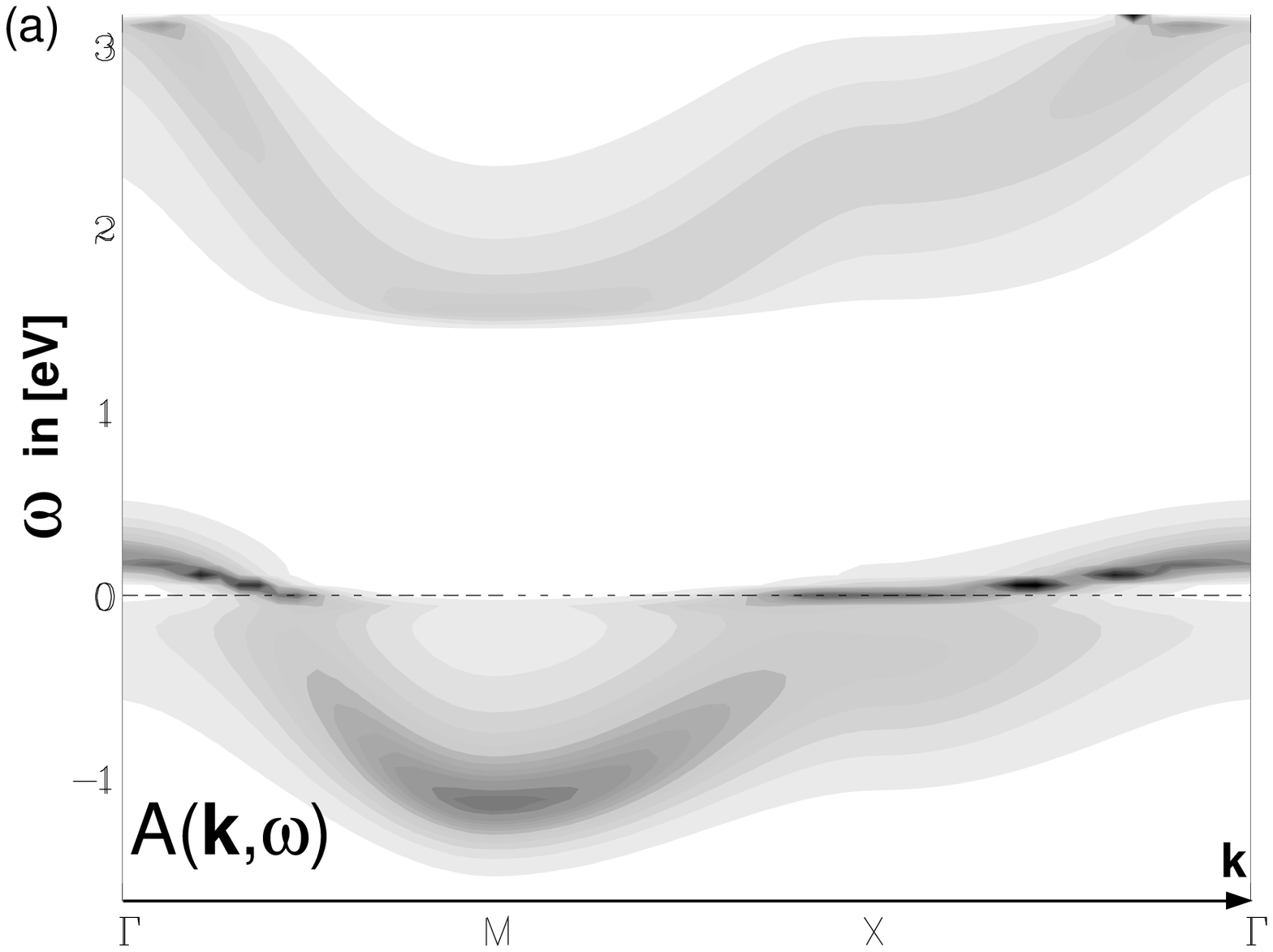}}}\\
\vspace{1.5cm}
\centering{\resizebox{0.4\textwidth}{!}{\includegraphics{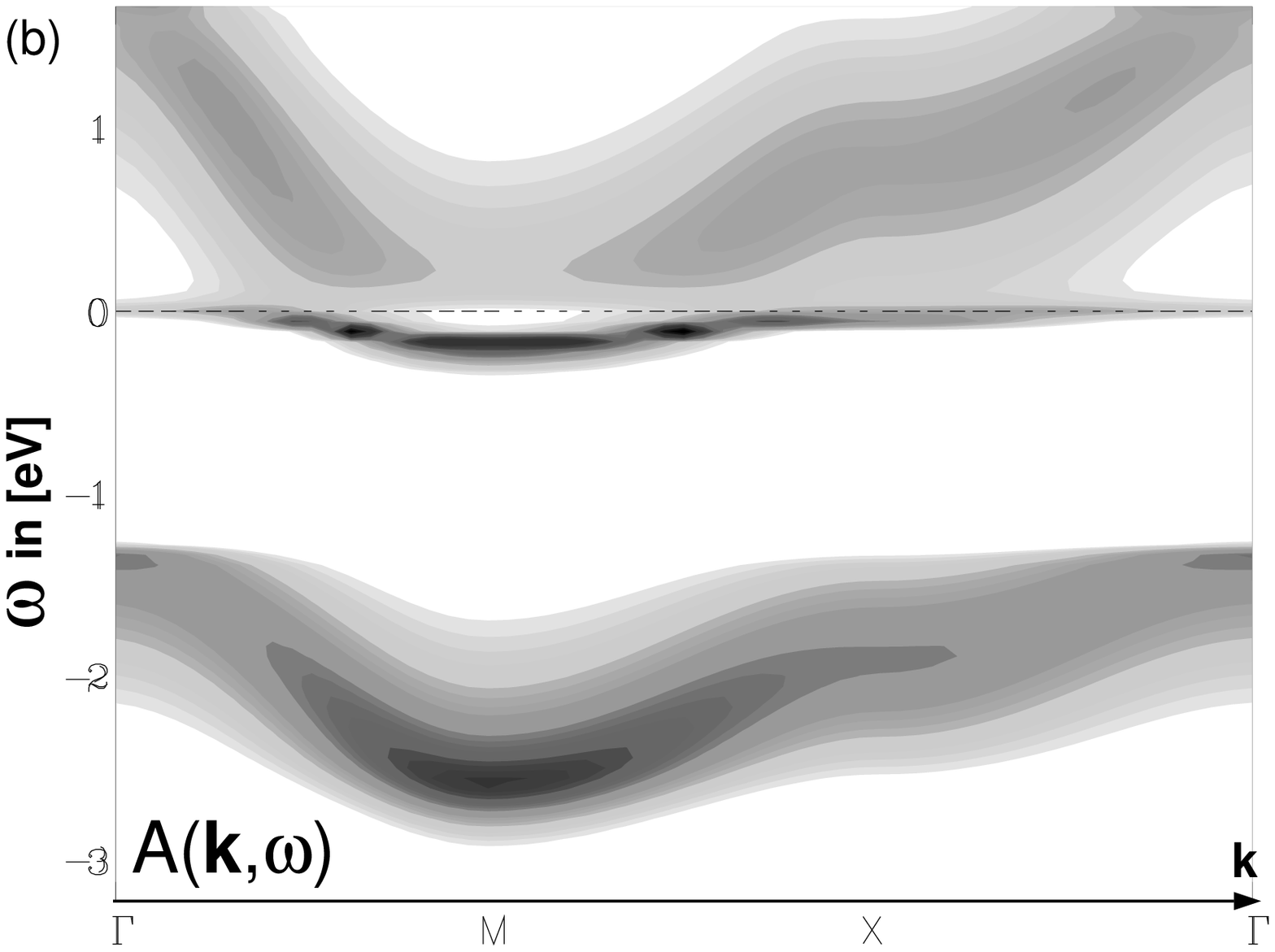}}}
\caption{Total spectral weight in a density-plot in the 
$\omega$-$\vec{k}$-plane for special $\vec{k}$-points for 
$\beta=40\:eV^{-1}$, $t=1\:eV$, $t^\prime=0.3\:eV$, $\Delta=3.6\:eV$, 
$U_d=7.2\: eV$ and $x=-8.5\%$ (a) , $x=+8.5\%$ (b).}
\label{fig:7}  
\end{figure}

In the following we want to argue that in order to solve the puzzle
of the correct choice of the parameter 
$t^\prime$, one has to consider electronic correlations.
In order to make contact with the free bandstructure we plot the total
spectral weight $A(\vec{k},\omega)$ in a density-plot in the
$\omega$-$\vec{k}$-plane. The dark regions refer to high spectral weight. 
In figure \ref{fig:7}a and \ref{fig:7}b the low energy part of the resulting 
$\vec{k}$-dependent total spectral weight for the electron and hole
doped $\rm CuO_2$-plane is shown.
These plots show the lower Hubbard band, the Abrikosov-Suhl resonance and 
the Zhang-Rice band. These bands exhibit the typical dispersion of the 
tight-binding $d$-band. 
On the other hand the $p$-bands and the upper Hubbard band at higher energies
(not shown in the figure) have the dispersion of their uncorrelated pendants.
For both types of doping there occurs a quasi-particle Abrikosov-Suhl 
resonance at the Fermi energy. It is important to note, that the dispersion 
of this band (which has $d$-character) completly determines the shape of the 
Fermi surface.   
\begin{figure}[ht!]
\centering{\resizebox{0.275\textwidth}{!}
{\includegraphics{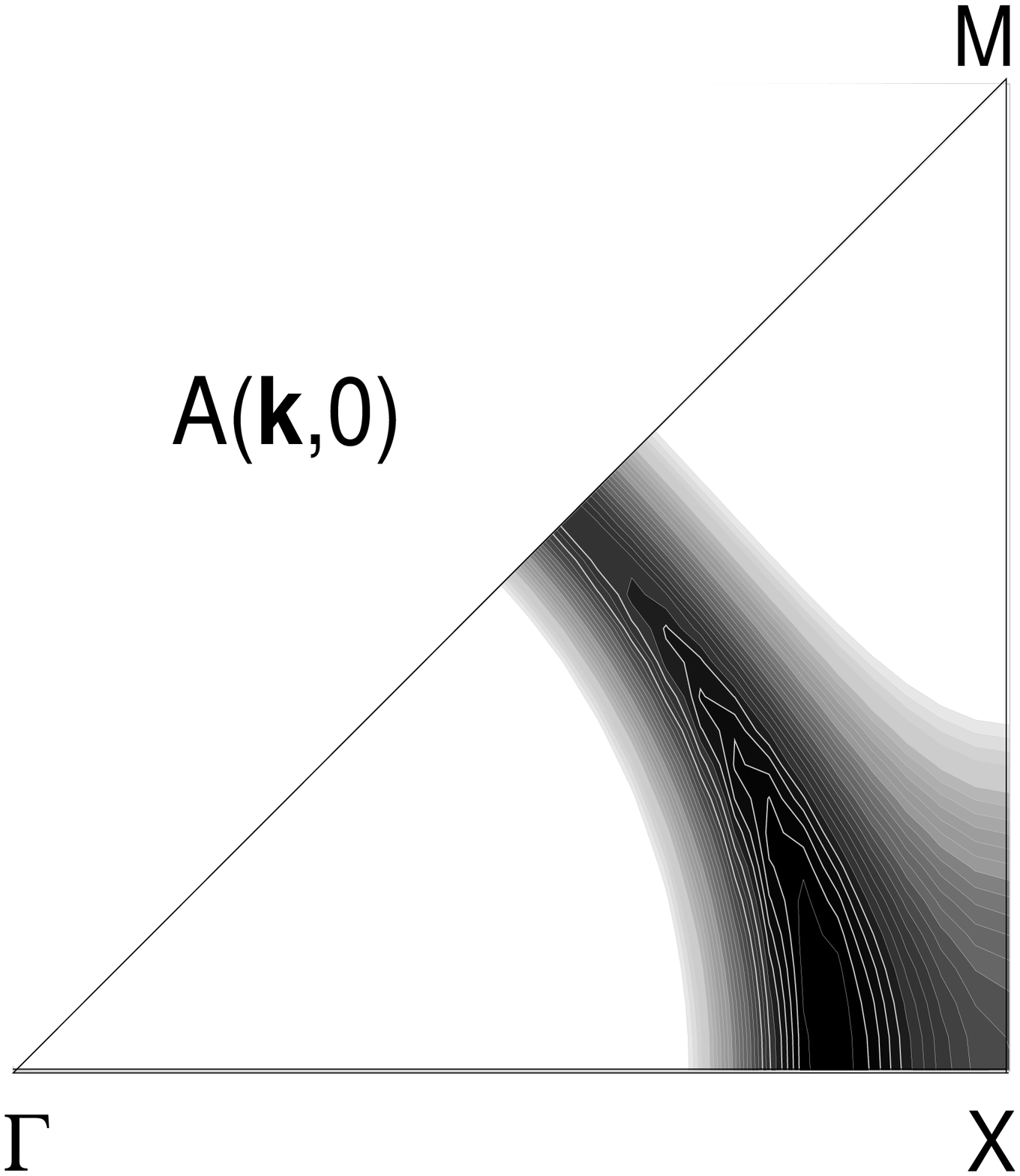}}}\\
(a) $t^\prime=-0.2t$\\
\centering{\resizebox{0.275\textwidth}{!}
{\includegraphics{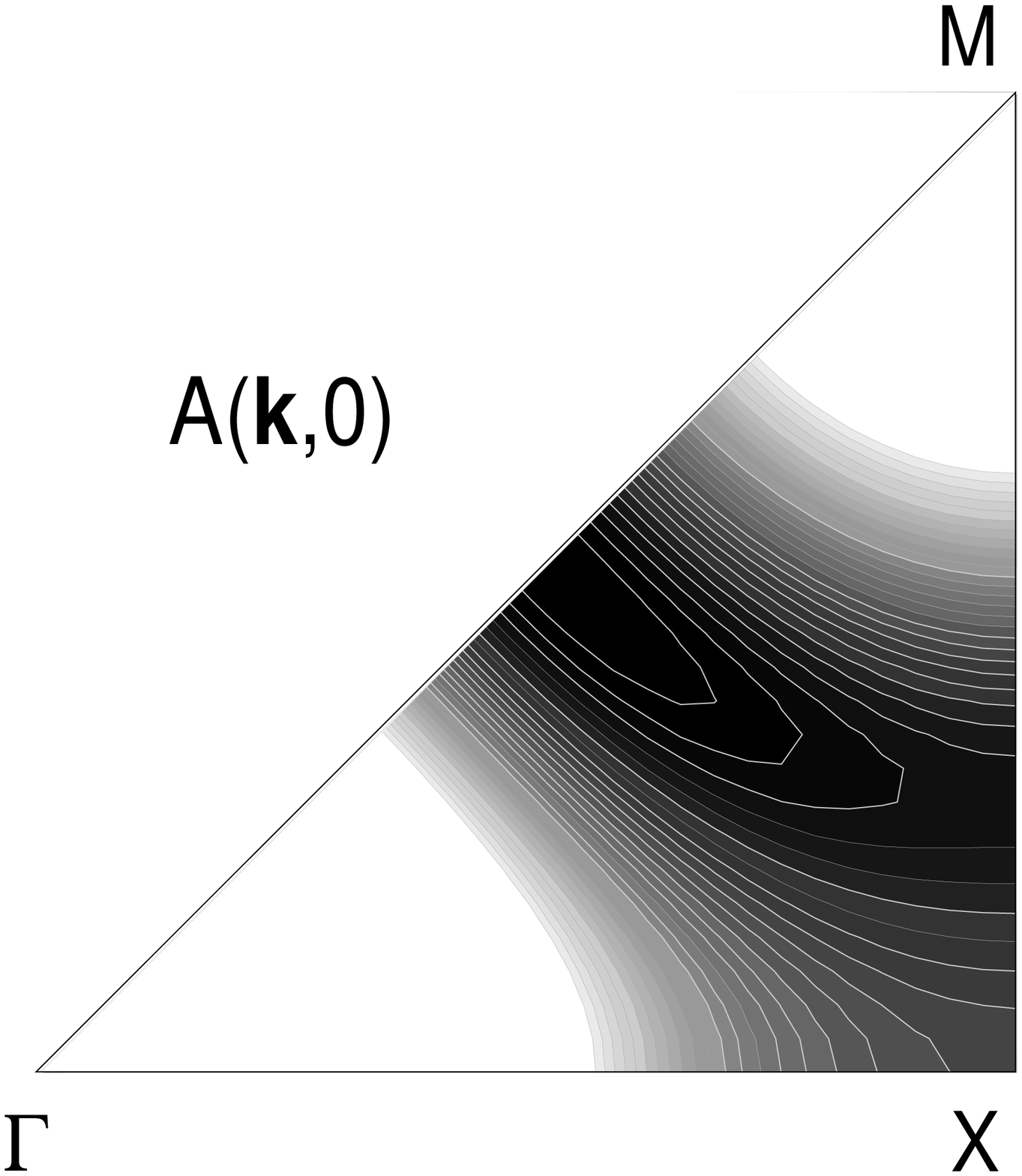}}}\\
(b) $t^\prime=+0.2t$\\
\caption{ Total spectral weight at the Fermi energy in a density-plot
  for two values of $t^\prime/t$ in the irreducible 
 part  of the Brillouin zone. The doping is $x=+5\%$ in both cases,
 the remaining parameters are the same as in figure \ref{fig:7}.
 One can see that the choice of $t^\prime/t$ determines the shape of the Fermi
 surface.}
\label{fig:9}
\end{figure}
In figure \ref{fig:9} we show, that for constant doping parameter $x=+5\%$,
one can change the shape of the Fermi surface by varying the value of
$t^\prime/t$, since it directly influences the dispersion of the band.
From the experimental fact that the Fermi surface encloses the $M$-point
we are led to a value $t^\prime/t>0$, which is in agreement with those
anticipated for high-T$_c$ compounds such as $\rm Nd_{2-x}CeCuO_4$ \cite{King} 
or $\rm Bi_2Sr_2CaCu_2O_{8+x}$ \cite{Dessau}.
    
\subsection{Electronic Response to a frozen phonon excitation}
\label{sec:8}

In this section we investigate the response of the oxygen occupation
on a level splitting of the $p_x$- and $p_y$-states.
Such a level splitting may be realized by a different displacement in
z-direction of the $x$- and $y$-oxygen in the $\rm CuO_2$-plane.
For simplicity we treat a 
symmetric energy-level splitting, i.e. one has to modify equation (7) to
(9) with ${\ep}_{x/y}=\ep\pm\delta\ep$.
With this modification we can look at the static
density-density correlation function in linear response approximation:
\begin{equation}
\langle\langle\delta\hat{n}_{xy},\delta\hat{n}_{xy}\rangle\rangle_{(\omega=0)}
\equiv \frac
    {\delta \langle \hat{n}_{x}\rangle-\delta \langle \hat{n}_{y}\rangle  } 
    {\delta_{\varepsilon_p}}
\end{equation} 
This quantity can be used to examine the electron-phonon coupling
for the $B_{1g}$-phonon in high-T$_c$-compounds. Of course it is not
possible to study the dynamical aspects of the coupling, 
but it is a rough meassure for the renormalization of the
phonon-frequency due to this coupling.
  
\begin{figure}[ht!]
\centering{\resizebox{0.4\textwidth}{!}{\includegraphics{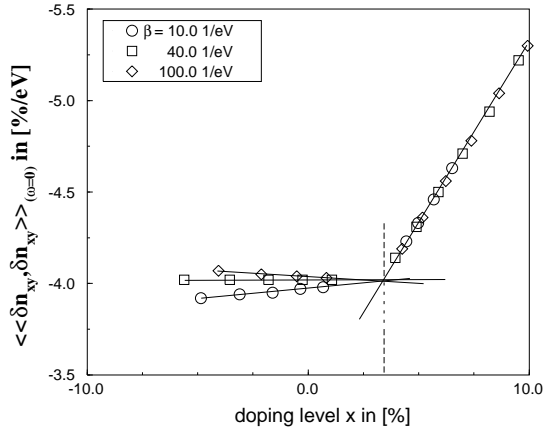}}}
\caption{Static density-density correlation at three
different reciprocal temperatures $\beta$ for $\delta\ep=0.1\:eV$, 
$t^\prime/t=0.5$, $U_d=7.2\:eV$ and $\Delta=3.6\:eV$.}
\label{fig:10}    
\end{figure}

Up to $\delta\ep=0.075\:eV$ the correlation function was found to be 
independent on $\delta\ep$ for all considered sets of parameters. 
The static susceptibility was therfore investigated for a constant value of
the level splitting, $\delta\ep=0.05\:eV$.
In figure \ref{fig:10} we show the density-density
correlation for $t^\prime/t=0.5$ for several
temperatures as a function of the doping $x$. 
The dashed line marks the doping which is found, when the chemical
potential $\mu$ is in the gap. 
Note that the crossing point is not exactly at $x=0$.
Thus the doping $x$ cannot be identified with the doping level of a real $\rm
CuO_2$-plane. 
This shortcoming originates from a numerical failure of the NCA and is
enhanced by a large value of $t^\prime/t=0.5$.
This effect can be neglected in the previous section \ref{sec:3}, 
where for $t^ \prime=0$ the model doping level $x$ for the insulating
case was less than $1 \%$. 
For clarity we term the region left to the dashed line in figure
\ref{fig:10} as electron
doping and the one right as hole doping.
In both regimes the density response is a linear function of the doping 
$x$. This holds for all investigated sets of parameters.
In the hole doped regime the density-density correlation shows in
addition no effect upon changing the temperature and it can
be described by a straight line with a slope
$\partial\langle\langle\delta\hat{n}_{xy},\delta\hat{n}_{xy}\rangle\rangle_{(\omega=0)}/\partial x\approx-6\:\:\:eV^{-1}$. 
For electron doped systems the slope 
increases with decreasing temperature.
\begin{figure}[ht!]
\centering{\resizebox{0.4\textwidth}{!}{\includegraphics{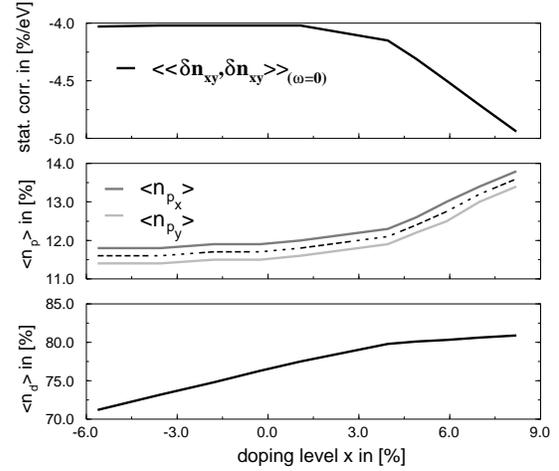}}}
\caption{Comparison of the static density-density correlation with the 
occupancies of the different orbitals for 
$\delta\ep=0.1\:eV$, $t^\prime/t=0.5$, $U_d=7.2\:eV$, $\Delta=3.6\:eV$ 
and $\beta=40\:eV^{-1}$.}
\label{fig:11}    
\end{figure}
In figure \ref{fig:11} the hole-density response of $p$-orbitals is
shown together with occupation of $p$- and $d$-states for different
doping. It can be seen that the desity response is strong in the hole
doped regime ($x>3\%$), when also the number of holes in the $p$-states
is large.

We have also calculated the hole-density response function for
$d$-states at $(\pi,\pi)$. This function is rather independent
of the Coulomb parameter $U_d$, but depends on the occupancy of the 
$d$-orbitals. 
For a calculation with $t^\prime=0$ the strength of the $d$-response varies
by a factor 3 to 4 compared to the $p$-response due to the larger
number of occupied $d$-orbitals. 

\section{Summary}
\label{sec:9}

In this paper we presented results for the electronic properties of
the three-band Hubbard model obtained with the help of the 
Dynamical Mean-Field Theory.
For the calculations we used two different model density of states:
a simple elliptic density of states to calculate the local spectral
density and the optical conductivity, a two dimensional
tight-binding density to calculate the details of the bandstructure
and a static response function.
 
The spectral density consists of a lower Hubbard band with
d-character, transitions into a Zhang-Rice band of p-d-singlet states, 
binding and non-binding p-bands, and a upper Hubbard band. For
electron doping the Fermi level is in the lower Hubbard band, for hole 
doping in the Zhang-Rice band. The energy separation between these two 
bands, which is the optical gap $\Delta_{opt}$ typically seen in
conductivity meassurements, is smaller than the bare  energy
separation of $p$- and $d$-states, $\Delta=\ep-\ed$. This tendency as
well as the calculated value of $\Delta_{opt}$ qualitatively agree
with experimental observed data.

At low energies the optical conductivity shows a narrow Drude-like
peak on a broad background as experimentally observed. The narrow
Drude peak  is due to transitions inside the Kondo-like feature close
to the Fermi energy.

We studied in detail the influence of an additional tigth-binding
coupling $t^\prime$ between oxygen states on the bandstructure.
It turned out that in the presence of correlations the variation of
the Fermi surface is considerably changed compared to the
non-interacting system. The most important aspect is that one finds
similar structures of the Fermi surface for both electron and hole
doped systems. Together with the experimental fact that the Fermi
surface encloses the $M$-point for both kind of doped systems we
conclude that $t^\prime/t$ must be positive.

We also investigated the response of the occupation-difference of
$p$-states on the energy splitting of the two different $p$-states in
the unit cell. Such a response function is a meassure for the
electron-phonon coupling of special phonons. Naturally this density
response increases with the occupation number of $p$-holes, i.e. it is 
larger if the Fermi energy is in the Zhang-Rice band than in the lower 
Hubbard band.

Acknowledgements: 
This work was partially supported by the DFG grant PR 289/5.
         
%
%
%
%


\end{document}